# Thermal decomposition and chemical vapor deposition: a comparative study of multi-layer growth of graphene on SiC(000-1)

D. Convertino[1,2], A. Rossi[1,2], V. Miseikis[1], V. Piazza[1], C. Coletti[1,3]
[1] Center for Nanotechnology Innovation @ NEST, Istituto Italiano di Tecnologia, Piazza San Silvestro 12, 56127 Pisa, Italy
[2] Laboratorio NEST – Scuola Normale Superiore, Piazza San Silvestro 12, 56127 Pisa, Italy
[3] Graphene Labs, Istituto Italiano di Tecnologia, Via Morego 30, 16163 Genova, Italy

## ABSTRACT

This work presents a comparison of the structural, chemical and electronic properties of multi-layer graphene grown on SiC(000-1) by using two different growth approaches: thermal decomposition and chemical vapor deposition (CVD). The topography of the samples was investigated by using atomic force microscopy (AFM), and scanning electron microscopy (SEM) was performed to examine the sample on a large scale. Raman spectroscopy was used to assess the crystallinity and electronic behavior of the multi-layer graphene and to estimate its thickness in a non-invasive way. While the crystallinity of the samples obtained with the two different approaches is comparable, our results indicate that the CVD method allows for a better thickness control of the grown graphene.

## INTRODUCTION

Epitaxial growth of graphene by thermal decomposition of silicon carbide (SiC) is a classical and successful approach to obtain large-area continuous films directly on a semi-insulating substrate [1,2]. In this process the SiC crystal acts as precursor: heating of the substrate at temperatures above 1300 °C in an Argon (Ar) atmosphere causes Si sublimation, and the C atoms left behind rearrange in a honeycomb structure forming one or more layers of graphene. Notably, the two different basal planes of the hexagonal polytypes of SiC, known as SiC(0001) (Si-face) and SiC(000-1) (C-face), show significantly different growth modes for graphene [1]. In particular, the graphene layers typically obtained on the C-face lack a defined azimuthal orientation (i.e., turbostratic graphene) so that each layer behaves as an isolated graphene layer and is electronically decoupled from the neighboring ones. For this reason, on this type of graphene, remarkable carrier mobilities have been measured [3]. However, control of the thickness of graphene obtained via thermal decomposition on the C-face of SiC is rather challenging. In previous work, Strupinski et al. [3] have shown that by adopting a classical chemical vapor deposition (CVD) approach, graphene with an appreciable thickness homogeneity can be obtained on SiC(0001). CVD on SiC(000-1) however, remains largely unexplored. In this study we compare the structural, chemical and electronic properties of graphene grown via thermal decomposition and via CVD on SiC(000-1).

## EXPERIMENTAL

Nominally on-axis 4H-SiC(000-1) substrates (Cree, Inc.) were first ultrasonically cleaned by acetone and isopropanol, subsequently immersed in diluted HF solution (10% HF) to remove the oxide layer and finally rinsed in DI water. In order to remove polishing scratches and to obtain atomically flat terraces the substrates were treated with hydrogen etching at a temperature of 1250 °C and a pressure of 450 mbar for 5 minutes [4]. Multi-layer graphene was subsequently grown on the C-face via thermal decomposition or via CVD. Hydrogen-etching and growth was carried out in a commercial resistively heated cold-wall reactor (Aixtron HT-BM). Thermal decomposition was achieved by heating the samples in Ar atmosphere at a temperature of 1350 °C and a pressure of 780 mbar for 15 minutes, similar to what described in [5,6]. CVD growth was achieved using methane ($CH_4$) as carbon precursor. Preliminary experiments were performed to optimize the growth process and it was found that addition of hydrogen in the gas mixture helped to reduce the formation of defects and to improve the homogeneity of the layer distribution. The sample was annealed at about 1350 °C within a 50% Ar and 50% $H_2$ atmosphere held at 780 mbar, while flowing 3 sccm of $CH_4$ for 5 minutes.

Analyses of the samples were carried out by using atomic force microscopy (AFM), scanning electron microscopy (SEM) and Raman spectroscopy. AFM examination was conducted by recording micrographs of different size (up to 20x20 µm) over several areas of each sample with an AFM+ from Anasys Instruments operated in tapping-mode. Large areas of each samples were inspected using a SEM (Merlin, Zeiss) operated in the in-lens mode. A micro-Raman spectroscope, equipped with a motorized sample stage (inVia Raman, Renishaw) and a 532 nm laser with a spot size of around 2 µm in diameter, was used to map the characteristic graphene Raman peaks.

**RESULTS AND DISCUSSION**

Figure 1 shows the typical morphology of hydrogen etched SiC(000-1) surfaces prior to graphene growth. No step bunching is observed, as the steps present unit cell height (i.e., 1 nm).

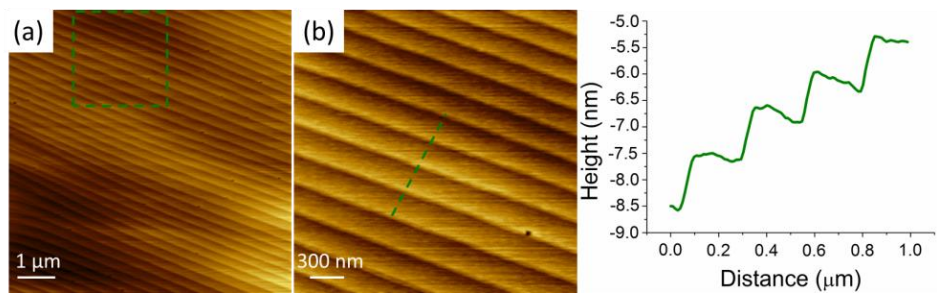

Figure 1. (a) Atomic force microscopy (AFM) of a 4H-SiC(000-1) wafer after hydrogen-etching. (b) zoom-in area with the relative height profile.

After the thermal decomposition process, AFM analyses revealed a surface topography similar to that reported in previous works [7,8]. As shown in Fig. 2(a) and in the related line profile, pleats (or ridges) 2 to 10 nm high are observed. Such features are typically retrieved on graphene on SiC(000-1) (whereas they are not found on epitaxial graphene on SiC(0001)) and are due to the differences in expansion coefficients of SiC and graphene and to their weak coupling [7]. Most notably, the surface displays a significant step bunching as visible in the representative AFM micrograph reported in panel (a). The related line profile shows a step with a height of about 40 nm. Adjacent micron-sized domains with different heights are also evidenced by the different grayscale contrast in the SEM micrograph in panel (b) and are similar to those described previously [9,10]. Indeed, the thermal decomposition process appears to take place in a rather quick and uncontrolled fashion, thus yielding a relatively inhomogeneous surface. Although the use of atmospheric pressure yields improvements with respect to the submicrometers domains typically obtained with UHV processes, the low surface energy of the C-face and the lack of a "growth enclosure" [11] most likely contribute to the obtained *canyon-like* morphology.

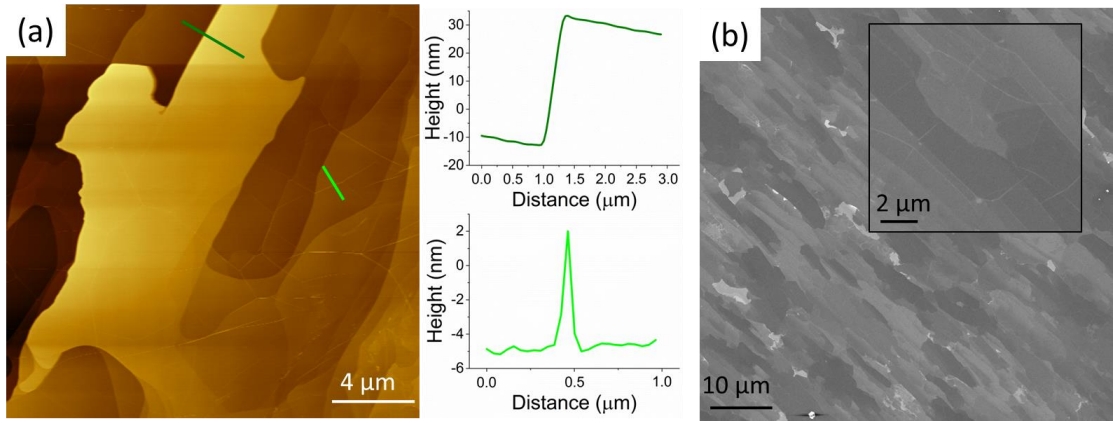

Figure 2. (a) AFM topography. Inset: profile analysis of a big step (top) and a ridge (bottom). (b) SEM image of graphene grown by thermal decomposition. Inset: high magnification of the same sample.

Graphene thickness was estimated via Raman by measuring the attenuation of the representative SiC peak measured at ~ 1516 cm$^{-1}$ [12]. This peak is an overtone of the L point optical phonon [13] and is attenuated in intensity when graphene overlayers are grown [12]. The remaining fraction of the substrate intensity after graphene growth is indicated as S and calculated as the multiplication needed to normalize the selected SiC peak in the bare SiC substrate to that of the graphene sample. The thickness ($t$) of the graphene overlayers is calculated as $t = -\ln(S)/2\alpha$ where $\alpha$ is graphene's absorption coefficient. Figure 3 reports micro-Raman maps of the ratio S (a) and of the 2D-peak full width at half maximum (FWHM) (b) of the same area of a typical sample. Values of S close to unity indicate a low attenuation and are therefore indicative of few-layer graphene. Instead, S values closer to zero are found for areas with thicker multilayer graphene. In the map in Figure 3(a) the fractional SiC Raman signal S ranges from 0.9 to 0.4, thus indicating that the number of layers varies between 3 and 23. The 2D peak can be fitted with one Lorentzian, indicating a non-Bernal-stacked structure [14], with FHWM ranging from ~25 cm$^{-1}$ to ~90 cm$^{-1}$. Areas with higher attenuation of SiC (lower values of S in Figure 3(a)), due to an increased thickness of graphene, correlate well with wider 2D FWHM as previously reported in [14]. A characteristic spectrum after the subtraction of a reference SiC is shown in panel (c). The G peak was found at ~1580-1585 cm$^{-1}$ and the 2D peak at ~ 2690-2720 cm$^{-1}$, both position ranges are comparable to those reported for similar samples in previous works [3,14]. Furthermore, absence of the disorder-induced D peak indicates high crystallinity of the grown graphene. However, despite the good crystallinity, such graphene samples might not be suitable for implementing devices. Indeed, if the graphene is to be used as an active material in (opto)electronic devices, significant differences in thickness and a *canyon-like* morphology are not ideal for their patterning and implementation.

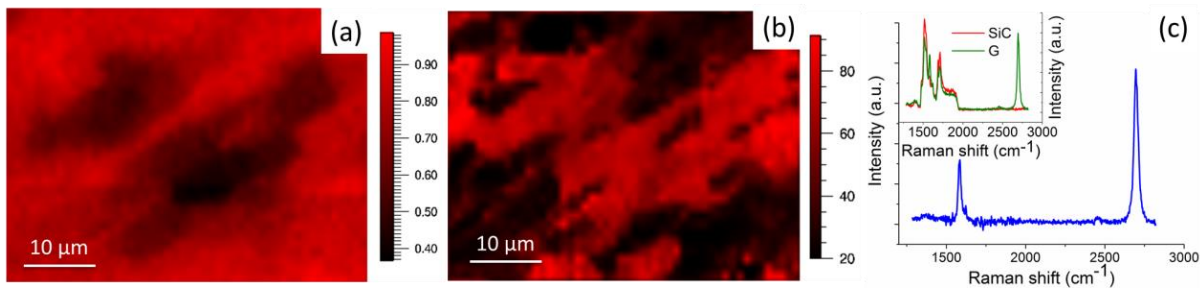

Figure 3. (a) fractional SiC Raman signal (S) and (b) 2D FWHM Raman mapping. (c) Raman spectra of graphene after SiC signal subtraction (blue color). Inset: original spectrum without subtraction (green colour) and SiC reference signal (red color).

As introduced before, another way to obtain graphene on SiC(000-1), is chemical vapor deposition by methane cracking. AFM analyses of the CVD-grown multi-layer graphene is reported in Figure 4(a). Notably, atomic steps are preserved as indicated by the relative line profile. Differently from thermal decomposition samples, the step-flow morphology is well-preserved and no dramatic variation in height within adjacent areas is observed. Locally, one can appreciate inclusions of domains with an increased roughness (although still in the few-nanometers range). The SEM micrograph in Figure 4(b) shows such inclusions as areas with a darker grayscale contrast. They most likely arise by a non-perfected CVD growth yielding the formation of sub-micrometer sized single-crystal domains. The grain size could possibly be increased by using higher hydrogen partial pressure during growth, an approach that needs to be investigated further.

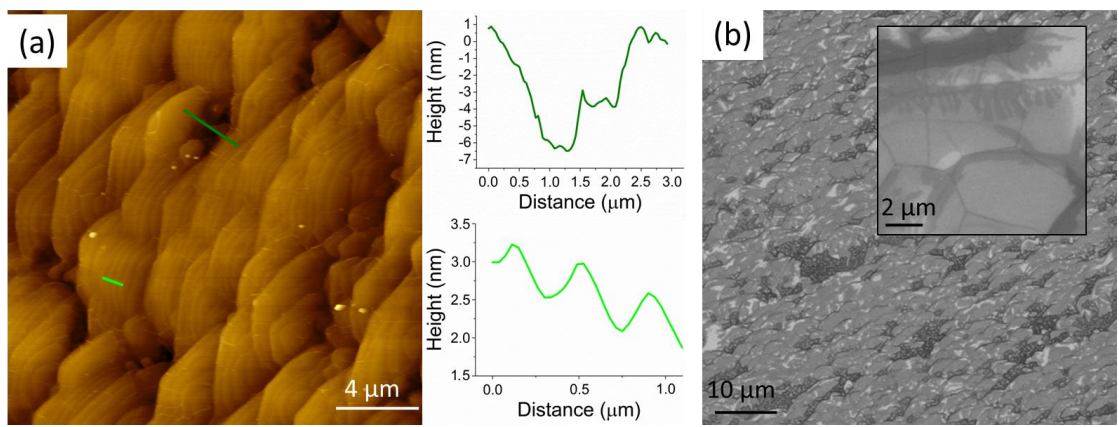

Figure 4. (a) AFM topography. Right insets: profile analysis of a rough area (top) and atomic steps (bottom). (b) SEM image of graphene grown by CVD. Inset: high magnification of the same sample.

Figure 5 shows Raman mapping of S (a) and of the FWHM of the 2D peak (b) for the same area of a typical CVD-grown sample. In this case, the value of S varies between 0.9 and 0.7, which indicates a much narrower graphene thickness distribution, ranging from 3 to 7 layers. As for thermal decomposition samples, the 2D peak can be fitted well with a single Lorentzian presenting a FWHM ranging from ~40 cm$^{-1}$ to ~60 cm$^{-1}$. A characteristic spectrum after the subtraction of a reference SiC is shown in panel (c). The G peak was found at ~1580-1595 cm$^{-1}$ and the 2D peak at ~ 2690-2720 cm$^{-1}$. The larger variation in the G peak position observed in the CVD samples might be indicative of graphene doping and needs to be investigated further. The D peak was generally absent and found only very occasionally, in spectra presenting a low 2D peak. This indicates that also in this case a high crystallinity is achieved.

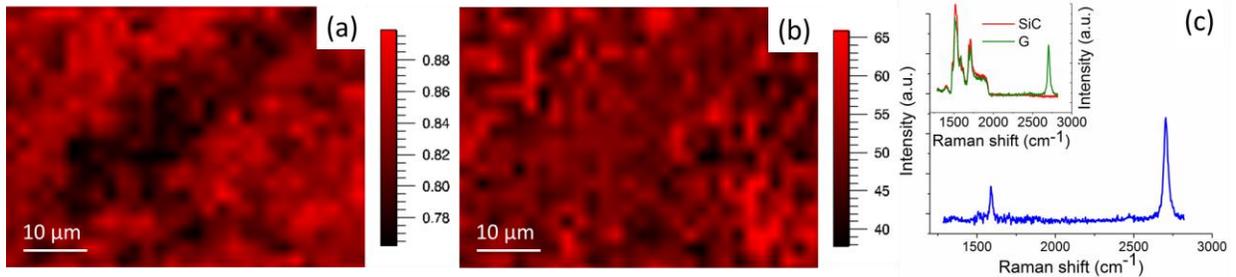

Figure 5. (a) fractional SiC Raman signal (S) and (b) 2D FWHM Raman mapping. (c) Raman spectra of graphene after SiC signal subtraction (blue color). Inset: original spectrum without subtraction (green colour) and SiC reference signal (red color).

By adopting a similar CVD process to that reported in this work, a high number of graphene layers (up to 90) with good crystallinity were synthesized, which were used for investigating absorption in the THz range [15]. As these samples were found to display saturable absorption, CVD-grown multi-layer graphene on SiC(000-1) might represent a favorable platform for the implementation of novel graphene-based mode-locked THz lasers [15].

**CONCLUSIONS**

In this work we compare the properties of multi-layer graphene grown on the C-terminated face of SiC by thermal decomposition and chemical vapor deposition. Raman analysis indicate that the graphene produced with both approaches present a high crystallinity and an electronic behavior typical of turbostratic samples, i.e. each graphene layer is electronically decoupled from the neighboring ones. AFM, SEM and Raman analysis show that the surface topography and the uniformity of the layer thickness distribution are improved significantly by using the CVD approach. By displaying appreciable thickness homogeneity and crystallinity multi-layer graphene grown via CVD on SiC(000-1) might represent an appealing platform for the development of novel optoelectronic applications.

**ACKNOWLEDGMENTS**

The research leading to these results has received funding from the European Union Seventh Framework Programme under grant agreement n°604391 Graphene Flagship.

We thank K. Teo and N. Rupesinghe from Aixtron for technical support with the Aixtron BM system.